\documentstyle{lamuphys}
\makeatletter
\let\chapter\hid@chapter
\makeatother
\def\etal{et~al.\ }
\begin{document}
\pagenumbering{arabic}
\title{The Nature of Compact Galaxies at z$\sim$0.2-1.3:
Implications for Galaxy Evolution and the Star Formation History of
the Universe.}

\author{R. Guzm\'an\inst{1}, A.C. Phillips\inst{1}, 
J. Gallego\inst{1,2}, D.C. Koo\inst{1} and J.D. Lowenthal\inst{1}}

\institute{UCO/Lick Observatory, University of California, Santa Cruz,
CA 95064, USA 
\and
Departamento de Astronom\'\i a, Universidad Complutense, 28040 Madrid,
Spain}

\titlerunning{Compact Galaxies at $z\sim0.2-1.3$}
\maketitle

\begin{abstract} We study the global scaling-laws of 51 compact field
galaxies with redshifts $z\sim 0.2- 1.3$ and apparent magnitudes
I$_{814}<23.74$ in the flanking fields of the Hubble Deep Field.
Roughly 60\% of the 45 compact emission-line galaxies have sizes, surface
brightnesses, luminosities, velocity widths, excitations, star formation rates
(SFR), and
mass-to-light ratios characteristic of young star-forming HII
galaxies. The remaining 40\% form a more heterogeneous class of
evolved starbursts, similar to local  disk starburst galaxies.  Without
additional star formation, HII-like distant compacts will most likely
fade to resemble today's spheroidal galaxies such as NGC 205. Our
sample implies a {\sl lower limit} for the global comoving 
SFR density of $\sim$0.004 M$_\odot$ yr$^{-1}$
Mpc$^{-3}$ at $z =$ 0.55, and $\sim$0.008 M$_\odot$ yr$^{-1}$
Mpc$^{-3}$ at $z =$ 0.85.  These values, when compared to a {\sl
similar} sample of local galaxies, support a history of the universe
in which the SFR density declines by a factor $\sim$10 from $z =1$ to
today. From the comparison with the SFR densities derived from
previous data sets, we conclude that compact emission-line galaxies,
though only $\sim$20\% of the general field population, may contribute
as much as $\sim$45\% to the global SFR of the universe at $0.4 < z <
1$.

\end{abstract} %

\section{Introduction} Galaxies exhibit a wide variety of correlations
among global parameters (such as luminosity, size, surface brightness,
velocity dispersion, colors, and line strength indices). These
empirical scaling-laws have been widely used to constrain
current theories of galaxy formation, and to measure
distances and peculiar velocities of galaxies to map the large-scale
distribution of matter in the nearby universe.  With the advent of the
Hubble Space Telescope (HST) and the new generation of 10-m class
telescopes, it is now possible to extend the study of
scaling-laws to galaxies at high redshift (e.g., Koo \etal 1995;
Van Dokkum \& Franx 1996;  Bender \etal 1996; Guzm\'an \etal 1996; 
Vogt \etal 1996). 
These new studies are proving key in our understanding of one of the major
unresolved questions in modern cosmology: how galaxies evolve with
look-back time.

	One of the most controversial
issues related to this question is the nature of the numerous faint blue
galaxies observed in deep images of the sky (see reviews by Koo 1996,
Ellis 1996, and references therein). The high surface density and weak clustering
of these galaxies argue against their being either the progenitors or
the merging components of present-day bright galaxies (Lilly \etal 
1991; Efstathiou \etal 1991).  Various theoretical
scenarios have instead suggested that the faint blue galaxies are
low-mass stellar systems experiencing their initial starburst at
redshifts $z \le 1$, some of which turn into the present population of
spheroidal galaxies (Sph), such as NGC 205 (Babul \& Ferguson 1996).  Given
their likely starburst nature, faint blue galaxies may also be major
contributors to the global star formation rate (SFR) density already
found to increase with lookback time to at least redshift $z \sim 1$
(Cowie et al. 1995, Lilly \etal 1996).

	In this project we investigate the ideas above on the
nature of the faint blue galaxies by comparing the scaling-laws of
distant low-mass starbursts to those of nearby galaxies. The goals
are: to identify their local counterparts, to assess their evolution
with look-back time, and to study their role on the star formation
history of the universe. A full description of the results summarized 
here can be found in Koo \etal (1995), 
Guzm\'an \etal (1996,1997), and Phillips \etal (1997).

\section{The Data} The galaxy sample consists of 51 compact galaxies selected
from I$_{814}$ HST images of the flanking fields around the Hubble
Deep Field (HDF; Williams \etal 1996).  These objects are compact in
the sense that they have small apparent half-light radii ($r_{1/2} \le
0.5$ arcsec) and high surface brightnesses ($\mu_{I814} \le 22.2$ mag
arcsec$^{-2}$). With no color information, the ``compactness'' criterion
optimizes the selection of dwarf stellar systems which are likely to be
low-mass starbursts. Spectra for these objects were obtained using LRIS at
the Keck telescope with a slitwidth of 1.1 arcsec and a 600 l/mm
grating.  The effective resolution is $\sim$3.1 \AA\ FWHM. 
Typical exposure times were 3000s.
The total spectral range is $\sim$4000-9000 \AA .  In
addition, we obtained two 300s direct V-band exposures with LRIS in
order to provide some color information.  Our
final data set includes: redshifts,
$V_{606}-I_{814}$ colors, absolute blue magnitudes ($M_B$), half-light
radii ($R_e$), surface brightnesses ($SB_e$), velocity widths
($\sigma$), masses ($M$), mass-to-light ratios ($M/L$), 
O[III]/H$\beta$ line ratios, and SFRs. 

	Of the 51 galaxies, 6 (or 12\%) show absorption-line spectra
characteristic of elliptical and S0 galaxies, while the remaining 45
(88\%) exhibit prominent oxygen and/or Balmer emission lines
and blue continua characteristic of vigorous star-forming systems or
narrow-line active galaxies. Most of the emission-line objects
are very blue with nearly constant $V_{606}-I_{814} \sim 0.9$, while
those with early-type spectra form a reasonably tight red sequence
just blueward of the color track expected for non-evolving elliptical
galaxies (Figure 1).  Hereafter we focus our
study on the emission-line compact galaxies. For convenience, we
divide this sample into intermediate- ($z<0.7$) and high-redshift
($z>0.7$) samples.

\section{Scaling-Laws} 
\begin{figure} 
\vspace{5.5cm}
\includegraphics{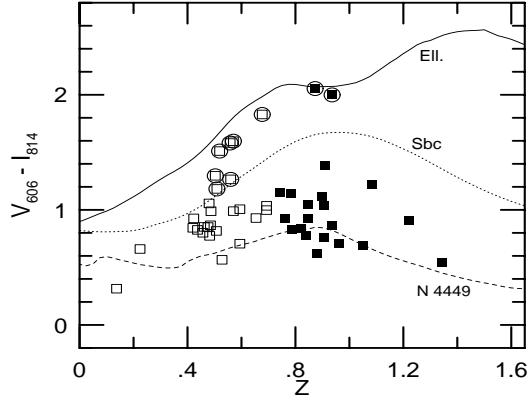}
\caption{$V_{606}-I_{814}$ color as a function of
redshift. The tracks of local elliptical, Sbc and starburst (NGC 4449)
galaxies are based on observed SEDs (no color evolution). Objects with
absorption-line dominated spectra are marked with ({\large $\circ$}).}  
\end{figure}
The global structural properties of galaxies can be adequately
described using the $M_B-SB_e$ and $R_e-\sigma$ diagrams. In these
diagrams, various galaxy types define distinct correlations, albeit
with large scatter (Figures 2a and 2b). Most of our sample galaxies
follow the sequence defined by young starbursts, such as
nearby HII galaxies or distant compact narrow emission-line galaxies
(CNELGs). To parametrize the starburst properties of the compact galaxy
sample we use the $[OIII]/H\beta-M_B$ and the $M-SFR/M$ diagrams
(Figures 3a and 3b). These diagrams discriminate among various types of
starburst and active galaxies. Most of the compacts with
$[OIII]/H\beta$ measurements lie in the moderate to high excitation
regime populated by HII galaxies and CNELGs. These objects also have
$SFR/M$ characteristic of HII galaxies. A second group of 
distant compacts have $[OIII]/H\beta$ and $SFR/M$ similar to those of 
more evolved disk starbursts such as local DANS and SBNs. 
Based on this, we have classified the sample into HII-like and disk 
starburst-like galaxies, depending on their $SFR/M$ 
(see Figure 3b). With this simple criterion, we find
that $\sim$60\% of distant compact galaxies have stellar  
{\sl and} structural properties consistent with those of nearby HII galaxies,
while the remaining 40\% are similar to more evolved disk starbursts.

\begin{figure} 
\includegraphics{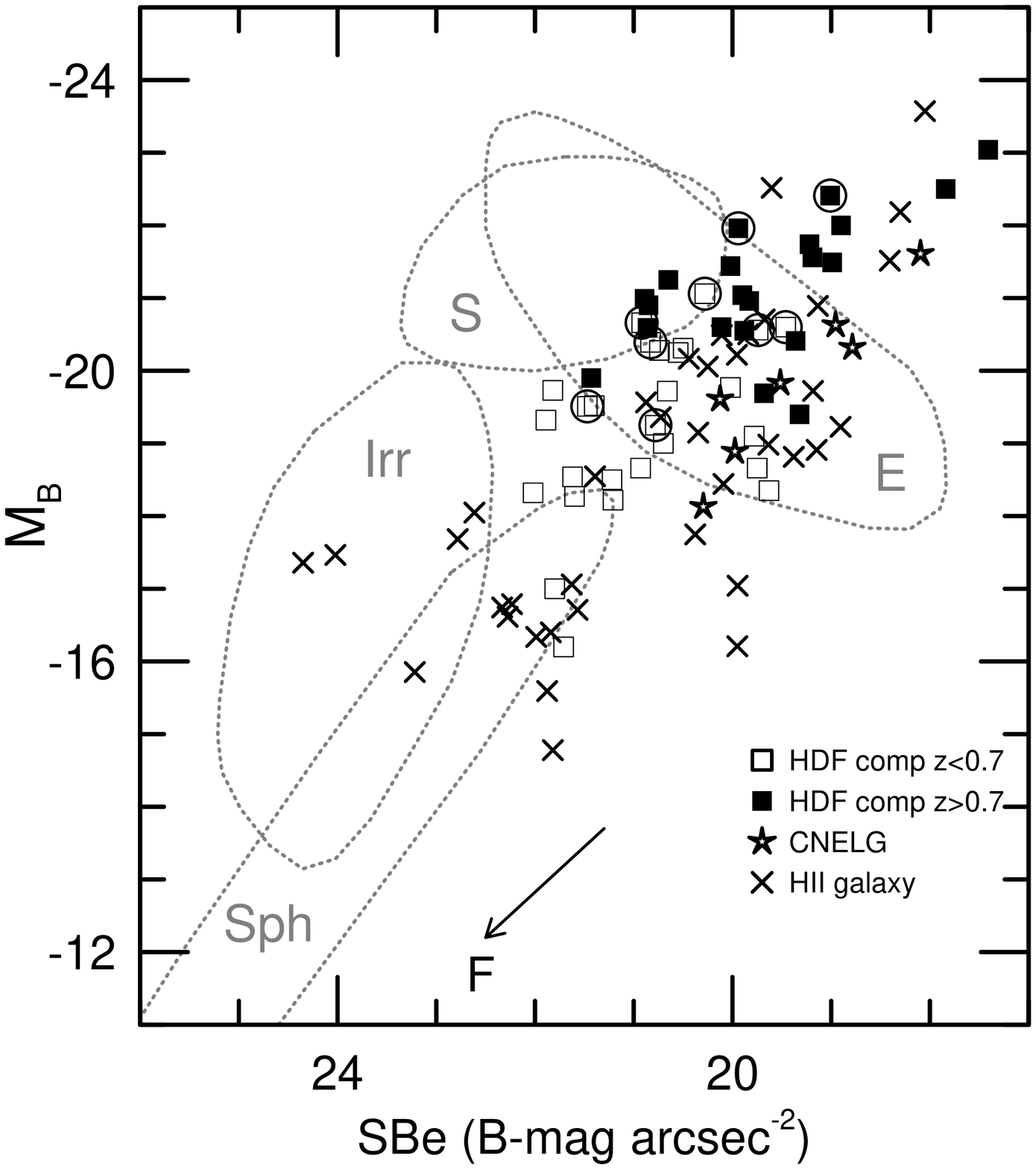}
\includegraphics{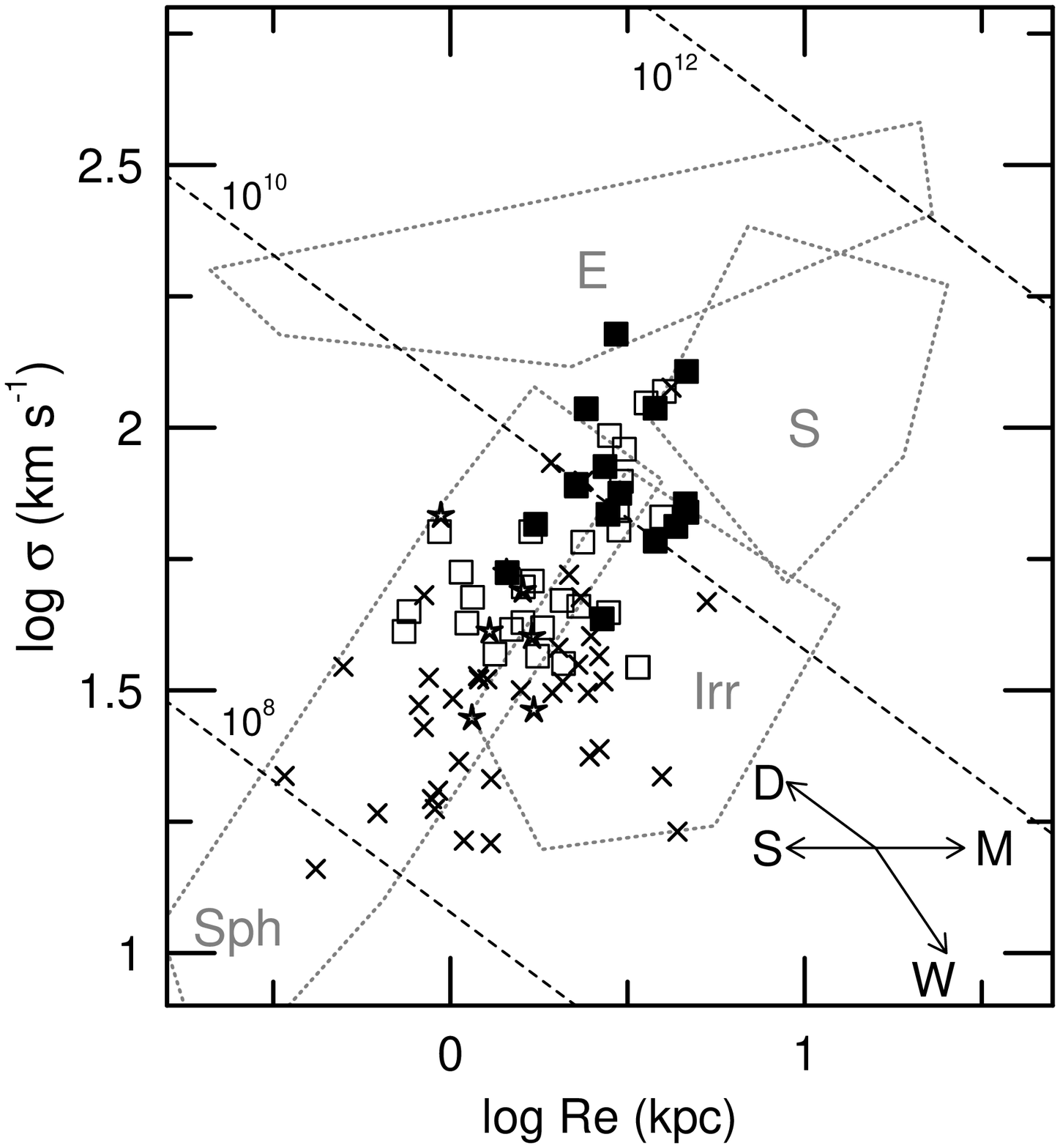}
\vspace{6.0cm} 
\caption{(a): $SB_e$ vs. $M_B$. Dotted lines 
indicate the general regions occupied by different classes of local 
galaxies; the arrow (F) represents the direction of fading. (b): 
$R_e$ vs. $\sigma$. Dashed lines represent constant mass-lines in 
$M_\odot$; the arrows represent the effect of dissipation (D), mergers (M),
stripping (S) and winds (W) on $R_e$ and $\sigma$. 
We adopt H$_0 =$ 50 km s$^{-1}$ Mpc$^{-1}$ and q$_0 =$ 0.05.
}
\includegraphics{figure3a.ps}
\includegraphics{figure3b.ps}
\vspace{7.0cm} 
\caption{(a): $M_B$ vs. $[OIII]/H\beta$. 
Local galaxy sample is from Gallego \etal (1997); DANS: Dwarf Amorphous 
Nuclear Starbursts; SBN: Starburst Nuclei; Sy2: Seyfert 2 galaxies; HII: 
HII galaxies. Dashed lines represent the approximate location of spiral 
galaxies. (b): $M$ vs. $SFR/M$. The dotted line represents the division 
between HII-like and disk starburst-like galaxies adopted in our
classification.}  
\end{figure}

	 Two other major results can be
drawn from the analysis of the $M-SFR/M$ diagram.  First, the highest
values of the $SFR/M$ exhibited by compact galaxies are similar 
to those of local HII galaxies. Thus we do not find
evidence for an increase in the peak of the $SFR/M$ activity with
redshift in our sample. Second, compact galaxies at $z>0.7$ are, on
average, $\sim$10 times more massive than their counterparts with
similar $SFR/M$ at $z<0.7$. Although selection effects may account for
the lack of low-mass compact objects at high-$z$, they cannot explain
why the massive star-forming systems are not present in the
intermediate-$z$ sample.  This is not the result of a volume-richness
effect either, since the volumes mapped in both samples differ by only a
factor $\sim1.5$. The apparent lack of massive starbursts in the
intermediate-$z$ sample suggests a steep evolution of the global SFR
with redshift, although this result should be taken with caution given
the small number of galaxies involved in this analysis.

\section{Discussion} 	

	Distant compact emission-line  galaxies are young, low-mass
star-forming systems.  Unless reignited by new star formation, they
should fade within a few Gyrs. The issue of fading and transformation
of one galaxy class to another is quite complex. Perhaps one of the
most useful tools we have to study how distant young galaxies relate
to nearby evolved stellar systems is the $R_e-\sigma$ diagram, since
neither $R_e$ nor $\sigma$ depend strongly on the fading of the
stellar population. Although there are several physical processes that
may modify these parameters during galaxy evolution (see Figure 2b),
we find no evidence against the idea that HII-like compact galaxies
(most of those with $M<10^{10}$M$_\odot$) are related structurally and
kinematically to the nearby population of Sph and Irr galaxies. Their
evolution into one galaxy class or another may depend critically on
their ability to retain part of their interstellar medium in the
likely event of starburst-driven galactic winds. The extremely low
mass-to-light ratios of HII-like compacts (i.e., $M/L\sim$0.3 solar)
suggest that the kinetic energy supplied by the current starburst is
large enough, compared to their binding energy, to blow out most of
the gas, thus preventing future star formation.  Without additional
star formation, galaxy evolution models predict that these low-mass
starbursts will fade enough to match the low luminosities and surface
brightnesses of Sph galaxies (see Figure 2a).  
We thus conclude that a class of
HII-like, faint blue galaxies may actually be among the progenitors of
today's spheroidals.

	The compact galaxy sample is also useful to investigate the
role of low-mass starbursts on the evolution of the SFR density at
redshifts $z<1$.  In Figure 4, we show a current overall picture of
the evolution of the SFR density with redshift. The interpretation of
this figure should be approached with caution, given the likely
differences in the calibrations for the various SFR tracers,
incompleteness of the data sets, and uncertainties in the
models. Despite these caveats, most of the results summarized in this
figure imply that the total SFR density of the universe decreased by a
factor of $\sim$10 from $z\sim$1 to the present-day. Assuming our
sample is representative of the general population of compact
galaxies, we estimate that the total SFR densities associated to this
class are: 0.004 M$_\odot$ yr$^{-1}$ Mpc$^{-3}$ at z=0.55, and 0.008
M$_\odot$ yr$^{-1}$ Mpc$^{-3}$ at z=0.85. These values, when compared
to a {\sl similar} sample of local galaxies, support a similar decline
in the SFR density in the last $\sim$8 Gyrs.  From the comparison with
the SFR densities derived by Cowie \etal (1995), we conclude that
compact emission-line galaxies, though only $\sim$20\% of the general
field population, may contribute as much as $\sim$45\% to the global
SFR of the universe at $0.4 < z < 1$.

\begin{figure}
\vspace{6.5cm}
\includegraphics{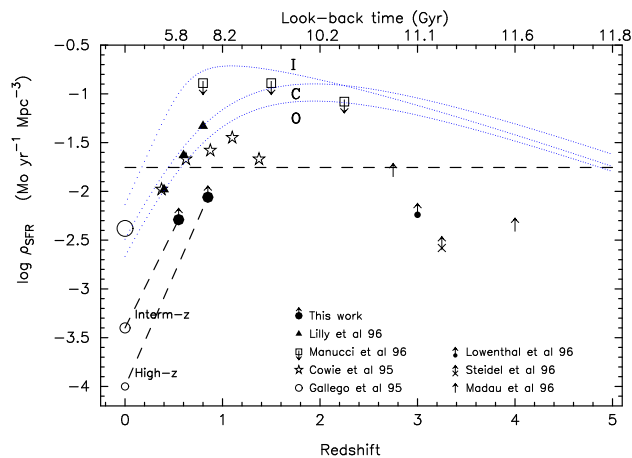}
\caption{SFR density vs. redshift. Filled circles are the estimates for 
compact galaxies. These values should be compared to the open circles
labelled ``Interm-$z$'' and ``High-$z$, which represent the values for
{\sl similar} samples of nearby compact galaxies. Dotted lines 
represent Pei \& Fall's models (1995). The dashed line represents the 
fiducial value. We adopt H$_0 =$ 50 km s$^{-1}$ Mpc$^{-1}$ and q$_0 =$ 0.5.}
\end{figure}

\vspace{0.5cm}
\noindent {\bf Acknowledgements.} This project is a collaborative effort
of the DEEP team at UC Santa Cruz 
(http://www.ucolick.org/$\sim$deep/home.html). R. Guzm\'an would like to
thank the organizing and scientific committees for their kind
invitation and financial support to participate in this excellent
meeting. Funding for this project is credited to NASA grants
AR-06337.08-94A, AR-06337.21-94A, GO-05994.01-94A, AR-5801.01-94A, and
AR-6402.01-95A from the Space Telescope Institute, and NSF grants AST
91-20005 and AST 95-29098. 

%
%

\end{document}